# *Network Traffic Anomaly Detection Method Based on Multi scale Residual Feature*

**Xueyuan Duan**

Department of Information Security,

Naval University of Engineering,

Wuhan, Hubei, 430033, China

and

College of Computer and Information Technology,

Xinyang Normal University,

Xinyang, Henan, 464000, China

**Yu Fu \***

Department of Information Security,

Naval University of Engineering,

Wuhan, Hubei, 430033, China

Email: morningqueen@139.com

*Corresponding author

**Kun Wang**

Department of Information Security,

Naval University of Engineering,

Wuhan, Hubei, 430033, China

and

School of Mathematics and Information Engineering,

Xinyang Vocational and Technical College,

Xinyang, Henan, 464000, China

**Abstract:** To address the problem that traditional network traffic anomaly detection algorithms do not sufficiently mine potential features in long time domain, an anomaly detection method based on multi-scale residual features of network traffic is proposed. The original traffic is divided into subsequences of different time spans using sliding windows, and each subsequence is decomposed and reconstructed into data sequences of different levels using wavelet transform technique; the stacked autoencoder (SAE) constructs similar feature space using normal network traffic, and generates reconstructed error vector using the difference between reconstructed samples and input samples in the similar feature space; the multi-path residual group is used to learn reconstructed error The traffic classification is completed by a lightweight classifier. The experimental results show that the detection performance of the proposed method for anomalous network traffic is significantly improved compared with traditional methods; it confirms that the longer time span and more S transformation scales have positive effects on discovering potential diversity information in the original network traffic.

## I INTRODUCTION

With the rapid development of science and technology, the Internet has been extended to various fields of society. Rich network applications have promoted the rapid development of society and economy. While changing people's lifestyle, they have also brought unprecedented challenges to network security [1]. Due to the openness of network protocols, Trojans, viruses and other malicious software are widely disseminated through the Internet, and various network attacks such as denial of service attacks and zero-day attacks against network protocols and application vulnerabilities have never stopped. These attacks not only affect the normal operation of cyberspace, cause great losses to the national economy, and even threaten national security.

Network traffic carries the information of network operation state, network behavior and user data. It is the carrier of information transmission and interaction in the network. By analyzing the characteristics of network traffic data, we can find the abnormal situation in the network, and provide reference for sensing network situation and formulating corresponding preventive strategies. It is of great significance to increase network protection ability and maintain network security. The reasons for network traffic anomalies can be divided into two categories [2]. The first is the performance reason, which is usually caused by unreasonable network topology design, misuse of users, and equipment failure. The second is the security reason, which is mainly due to the abnormal network traffic caused by the criminals' intrusion attacks on the network by using security vulnerabilities, such as DOS (Denial of Service), U2R (User to Root) and other attacks. This paper mainly studies the network traffic anomaly caused by security reasons.

Network traffic anomaly detection is to use various detection technologies to analyze and judge network traffic and discover potential attacks in the network, which is an effective network security protection method. Previous scholars have tried to use machine learning methods such as K-Means[3], Naive Bayes [4][5], Support Vector Machine[6][7], Decision Tree[8] for anomaly detection of network traffic. However, with the continuous extension of the network boundary and the surge of various service applications, traffic data shows explosive growth. Traditional machine learning methods often face problems such as difficulty in feature design, high false alarm rate and weak generalization ability when dealing with massive, high-dimensional and dynamic network traffic.

Deep learning has a strong representation ability, which can automatically extract features from the original data, and is widely used in natural language processing, machine vision, intelligent diagnosis and other fields. Due to the strong temporal correlation of network attacks, when anomaly detection of network traffic is carried out, Recurrent Neural Network (RNN) [9] and Long Short-Term Memory (LSTM) [10] are often used to capture the implicit correlation in network traffic data, namely the potential connection between the current connection and the connection in the previous period. In addition, in view of the scarcity of some types of samples in traffic, in addition to the solution of "resampling" Generative Adversarial Network (GAN)[11] is usually used to generate a few types of samples to alleviate the problem of data imbalance. In addition, the error between the reconstructed sample and the real sample generated by GAN is often used as the basis for anomaly determination [12][13]. Studies have shown that the length of observation time interval, i.e., the time scale of the flow, is the key factor affecting the detection performance[14].

The above methods only analyze the correlation of features from the time domain perspective and do not utilize the frequency domain features of flow data. Some signal experts have found that the time-varying signal of network attack traffic is quite different from that of normal traffic in frequency characteristics[15]. Wang *et al*.[16] proposed a wavelet-based neural network structure that seamlessly embeds wavelet frequency analysis into deep learning framework to learn the features of network traffic by taking advantage of wavelet decomposition in frequency learning. a novel Channel Boosted and Residual learning based deep Convolutional Neural Network (CBR-CNN) architecture is proposed for the detection of network intrusions[17], which uses multiple Stacked Auto-Encoders (SAEs) for multiple mapping of the original signal to achieve channel boosting effect, and then uses the residual network to learn the features at different granularities in each channel. Although these two methods exploit the frequency features of the network signal, they are limited to a certain scale and do not fully exploit the different scale features of the network signal. For the classification task, the characteristics of high-frequency level can better reflect the difference in fine-grained flow data; for the prediction task, the characteristics of the low-frequency level mainly reflect the original state of the flow data, which is conducive to capturing the trend information to reflect the future trend. It can be seen that network traffic shows different behavioral characteristics at different time scales, and it can also reflect the original state and fine-grained difference of signals in different frequency domains.

Most current deep learning anomaly detection methods are single-layer architectures and do not fully utilize the feature information of network traffic at different time and frequency scales. Therefore, we propose a multi-scale residual feature-based anomaly detection method for network traffic. The proposed method uses sliding window and wavelet transform techniques to decompose and reconstruct the original network traffic at multiple levels and obtain reconstructed sequences on different frequency domains; the multi-scale reconstructed sequence samples of normal network traffic are used to model the SAE network and construct a similar feature space, which can generate samples similar to the normal traffic, and the reconstructed error vector between the generated samples and the input samples is used as the input of the residual network group, and the residual network group aggregates the captured diversity information and inputs it to the classifier to complete the classification prediction of the original input traffic. The research work in this paper is as follows:

(1) The generation method of multi-scale patterns of the original network traffic is designed, and the wavelet domain data of the original information at different time scales and frequencies are obtained.

(2) The SAE generation framework is constructed to model the SAE by using normal traffic samples, and the SAE learns that the real distribution of normal traffic has the ability to complete the reconstruction of normal traffic samples only.

(3) Similar Feature Space is constructed to map the input data features into reconstructed feature space for obtaining discriminating error feature.

(4) Based on the concepts of residual transformation and merging, a multi-path residual group network is constructed to capture diversity features of the input data.

(5) The performance of the proposed model is evaluated on several publicly available network traffic datasets. Experiments show that our proposed method is able to perform the anomaly determination task of network traffic and improves the metric performance over current typical anomaly detection models, and verifies that multi-scale information has a positive impact on anomaly detection.

This paper is organized as follows. In Section Ⅱ, some theories related to this study are introduced; in Section Ⅲ, the construction and functions of our proposed Multilevel Scale Residual Classifier (MSRC) are described; in Section Ⅳ, the requirements of the experimental environment setup and the experimental data are explained; in Section Ⅴ, three groups of experiments are designed to verify the feasibility of our method, and several typical anomaly detection methods are compared to evaluate the performance of our anomaly detection method; finally, a conclusion is given in Section Ⅵ.

## Ⅱ RELATED THEORY

### 2.1 Discrete Wavelet Transform and Reconstruction (DWT&R)

We know that large observation scale can show the global trend of signal data, while small scale pro-vides more details. Discrete Wavelet Transform ( DWT ) [18]can decompose the original signal into a set of wavelet basis functions $\{\psi_{a,b}(x)\}$; $\psi_{a,b}(x)$ is generated by mother wavelet $\psi(x)$ defined in a finite interval by translation and scaling

$$\psi_{a,b}(x) = \frac{1}{\sqrt{a}}\psi(\frac{x-b}{a}) \qquad (1)$$

where $a$ is the scaling parameter and $b$ is the translation parameter. They are multiples of $2^j$, where j > 0 and integer.

For the given original sequence $\mathrm{x} = \{x_1, x_2, ..., x_n\}$, the original sequence x is decomposed into an approximate part $\mathrm{x}^l$ (approximations) and a detailed part $\mathrm{x}^h$ (details) using Mallat's algorithm. The approximate part represents the low-frequency component of the original signal, containing more rough information than the original information, while the detail part is the high-frequency component of the original signal, containing the fine-grained information in the original signal. Multi-level decomposition of low-frequency components can be carried out by using the method of "down sampling", and more low-frequency components with low resolution can be obtained. This iterative process is shown in formula ( 2 ):

$$\mathrm{x}^l_{i-1} = c^h_i \mathrm{x}^h_i + c^l_i \mathrm{x}^l_i, \quad i \in N \qquad (2)$$

where $\mathrm{x}^l_0 = \mathrm{x}$, $c_l$ and $c_h$ are the approximate coefficients and detail coefficients of wavelet transform, respectively. The approximate coefficients are obtained by convolution of the original signal with the low-pass filter, while the detail coefficients are obtained by convolution of the high-pass filter with the original signal.

The multi-scale approximation and detail sections can reflect the rich information of the data at multiple levels, where the higher-level approximation section represents an overall trend behavior, while the detail section at each level can represent more local information. Different levels of layers have different time and frequency resolutions, and the frequency resolution increases and the time resolution decreases with increasing levels of layers.

The coefficient list $c = \left[ c^k_l, c^k_h, c^{k-1}_h, ..., c^2_h, c^1_h \right]$ can be generated after the original signal is transformed by $k$ layers of wavelet. Using this list of coefficients, the signal reconstruction at level $j$ can be completed as shown in formula (3):

$$R_j = f(\sum_{i=j}^{k} c^h_i \mathrm{x}^{h,\uparrow}_i + c^l_k \mathrm{x}^{l,\uparrow}_k), j \in [1,k] \qquad (3)$$

where $f(\bullet)$ represents the reconstruction function, $\mathrm{x}^{h,\uparrow}_i$ and $\mathrm{x}^{l,\uparrow}_k$ are obtained by "up-sampling" $\mathrm{x}^h_i$ and $\mathrm{x}^l_k$ respectively, and $k$ is the number of decomposition levels. The maximum number of levels of wavelet decomposition that can be performed for original data of length n is $\log_2^n$, so $k \in [1, \log_2^n]$. For the reconstruction of the data at level $j$, it can be synthesized by up-sampling the approximate and detailed solutions at level $j+1$ and convolving them using the reconstruction filter when $j$ takes different values so that we can transform the original data sequence into some data sequences of different scales. Figure 1 shows the multi-level discrete wavelet decomposition and reconstruction process.

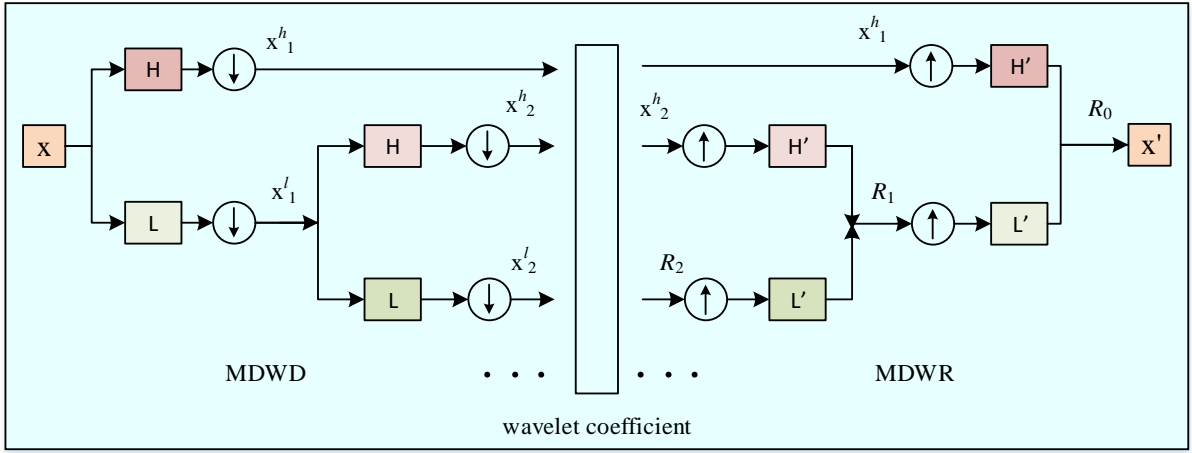

Fig.1: The framework of Multilevel Discrete Wavelet Transform

## 2.2 Stacked autoencoder (SAE)

AE is feed-forward neural network, which consists of input layer, hidden layer, and output layer, usually the input and output layers have the same size, by mapping the input to the latent feature space, and then mapping the latent space back to reconstruct the output, this process can be expressed as encoding and decoding[19]

$$\begin{aligned} z &= W_e x + b_e \\ h &= f(z) \\ x' &= \sigma(W_d h + b_d) \end{aligned} \quad (4)$$

where $x$ is the input, $h$ is the latent variable, and $x'$ is the reconstructed output of $x$ after encoding and decoding; $f(\cdot), \sigma(\cdot)$ are the activation functions; $W_e$、$W_d$、$b_e$、$b_d$ are the weights and biases of the encoder and decoder, which are optimized by minimizing the reconstruction error $\min(x - x')^2$.

SAE is a multilayer neural network stacked by multiple AEs, and the structure is shown in Fig. 2. The latent variables output from the former AE encoder are used as the input of the latter AE encoder, and this chain connection can help us capture more relevant details of the original feature space during the conversion process from the original feature vector to the destination feature vector, which helps to improve the performance of the classifier. Let $\phi_i$ and $\varphi_i$ be the encoder function and decoder function of the $ith$ AE, respectively, then the encoding process of SAE can be represented by the superposition of the conversion functions of each AE encoder as follows:

$$\begin{aligned} x_w = x_{encoded} &= \phi_w(\cdots(\phi_1(x))) \\ &= \phi_w \circ \phi_{w-1} \circ \cdots \circ \phi_2 \circ \phi_1(x) \end{aligned} \quad (5)$$

where $x_w$ denotes the higher-order feature representation of the original input $x$ output from the $wth$ activation function after a series of transformations. The decoding process then uses the decoder conversion function of each AE to reconstruct $x_w$ in reverse order, and the process can be expressed as:

$$\begin{aligned} x_{rec} = x_{decoded} &= \varphi_1(\cdots(\varphi_w(x_w))) \\ &= \varphi_1 \circ \varphi_2 \circ \cdots \circ \varphi_{w-1} \circ \varphi_w(x_w) \end{aligned} \quad (6)$$

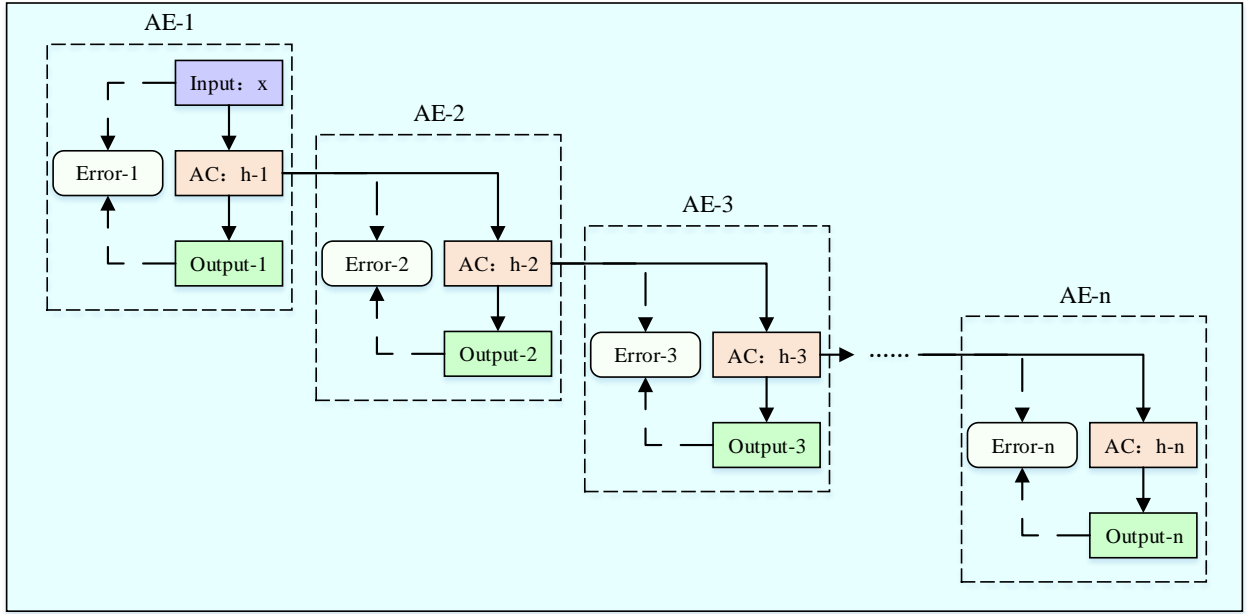

Fig.2: Structure of SAE

We adopt a greedy hierarchical training approach for SAE, including two stages of pre-training and fine-tuning.

Pre-training: Using an unsupervised method, the SAE is initially trained layer by layer using the mean square error of the minimization reconstruction error. The trained front layer takes the learned potential representations as the input of the back layer, trains the back layer to optimize the parameters, and later passes the learned new representations backward again to help the following layers to train until the training of all AE blocks is completed, and this pre-training approach can provide good initial parameters for the whole SAE.

Fine-tuning: Input normal data with labels, cross-entropy loss function and stochastic gradient descent optimization methods are used to precisely tune the parameters of SAE using back propagation. In this way, SAE is able to learn the distribution of normal samples and generate more accurate samples of normal data than abnormal samples.

SAE is trained only on normal data samples, and when data including both normal and abnormal samples are input to SAE, it can reconstruct the normal samples accurately, however, larger biases will occur in the reconstruction of anomalous samples.. The absolute difference between the original vector and the corresponding component of the reconstruction vector is taken as the reconstruction error vector, and it is used as the identification feature to identify normal and abnormal samples, and the reconstruction error vector can be expressed as:

$$e_i = |x_i - \hat{x}_i| \qquad (6)$$

Here, $x_i$ is the $i$-th component of the original feature vector, $\hat{x}_i$ is the reconstructed feature vector of the $i$-th component, and $e_i$ is the reconstructed error vector of that component.

2.3 Residual networks

Residual networks, Resnet[20] consists of a residual block (RB) based on the concept of residual learning. Originally designed to address the performance degradation of deep neural networks, it is mainly used in convolutional neural networks. This connection method breaks the traditional convention that the output of n layers of a neural network can only be given to n+1 layers as input, so that the later layers can learn the residuals directly and maintain the integrity of the information, while the whole network only needs to learn the part of the difference between input and output, simplifying the learning goal and difficulty. The structure of the residual block is shown in Figure 3, where Wight is the convolutional network layer, ReLU is the activation function, and $\oplus$ is the corresponding element operation.

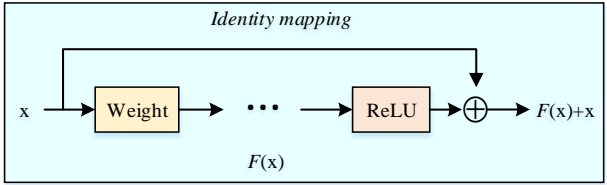

**Fig.3**: Residual block of Resnet

Assuming that we use the deep network expectation to complete the mapping $x \rightarrow H(x)$, then the expectation mapping $H(x)$ can be expressed as:

$$H(x) = F(x) + x \qquad (7)$$

At this point, the residual function can be obtained:

$$F(x) = H(x) - x \qquad (8)$$

It can be seen that $F(x)$ is the network mapping before computation $\oplus$, $H(x)$ is the network mapping after computation $\oplus$, and Identity mapping is equivalent to equal mapping. This residual learning structure of "shortcut connection" in which x bypasses the residual function $F(x)$ in the forward neural network can perform mapping at any optimal position without adding additional parameters or increasing the computational complexity of the network.

## III  MODEL

In this section, our proposed network traffic anomaly detection model with multi-scale residual features is presented in detail. The research work in this paper focuses on the following four aspects.

### 3.1 Multi-scale feature representation

As mentioned above, the original signal after discrete wavelet transform can produce more abundant data sequences, which can represent more local features in addition to the original ones. When large-scale network traffic is considered as a signal, it is found that the time-varying signal of normal network traffic has different frequency characteristics from the time-varying signal of abnormal traffic. Therefore, we will use the sliding window and wavelet transform methods to transform the original traffic data to get the reconstructed sequence data with different time span and frequency scales to explicitly learn the multi-scale information of the traffic. According to formula (3), we can get the wavelet decomposition and reconstructed data sequence as:

$$R_{j,sw} = f(\sum_{i=j}^{k} c_i^h x_{i,sw}^{h,\uparrow} + c_k^l x_{k,sw}^{l,\uparrow}), j \in [1, k] \qquad (9)$$

where $sw$ is the sliding window size, so $k \in [1, \log_2^{sw}]$.

### 3.2 Constructing similar feature space

We connect the pre-trained encoder one by one, and connect the corresponding decoder in the opposite order to form the SAE framework that can generate similar feature space, as shown in Fig. 4.

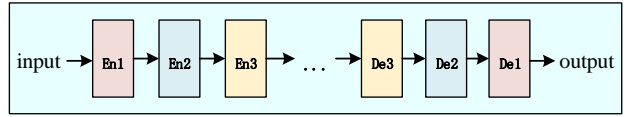

**Fig.4**: SAE for Similar Feature Space

SAE is modeled using normal network traffic data, so that SAE can fully learn the distribution space of normal traffic feature and construct the similar feature space. In the similar feature space, the reconstruction of normal data can only be completed, that is, the similar feature space can only generate samples similar to the normal feature space. For known or unknown anomalies, the samples generated by similar space reconstruction will show great difference from the original samples.

The reconstructed sequence data of the original flow at multiple scales generated in Section 3.1 are input into the corresponding SAE, and the elemental difference is made with the output after SAE mapping transformation to obtain the reconstructed error of the original flow at multiple scales.

## 3.3 Constructing multi-channel residual block groups

In order to alleviate the degradation of deep network performance and improve the diversity and controllability of network representation of features, a residual group structure based on parallel multiple residual blocks is proposed. In order to ensure the diversity of residual transformation, each residual block has different layers, so that the residual group can learn the diversity transformation from simple to complex. In addition, in order to ensure the controllability of residual transformation, only the first and the last parts are directly mapped on each residual path, and the connection is no longer opened in the middle network layer.

As shown in Fig. 5, the residual group (RG) with three residual blocks in parallel is shown, where Wight is the convolution operation in the convolution network, BN is the batch standardization, ReLU is the activation function, compose is the summary of the transformation results of each residual block, and FC is the operation of the full connection layer. The reconstructed error vector of each scale is output by SAE, and the data is mirrored and output to multiple residual blocks. The results obtained by residual block transformation are aggregated and output after full connection layer.

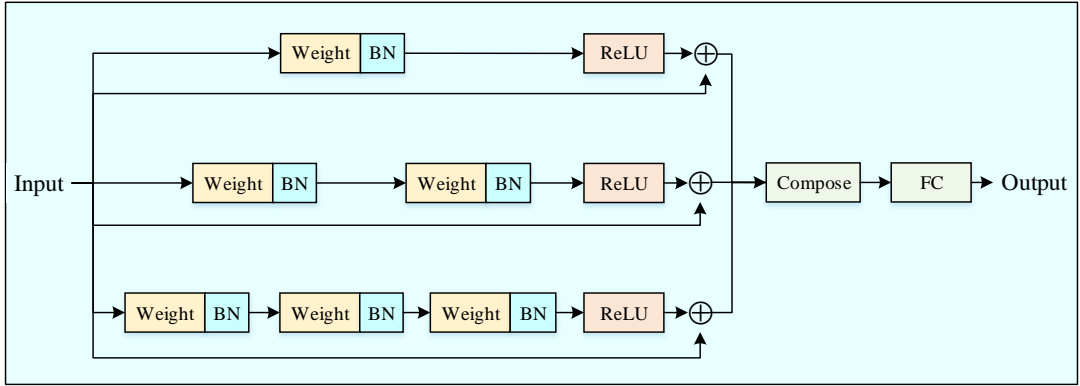

**Fig.5:** 3-level residual group structure diagram

Assume the input is x, Complete the transformation in the $i$th residual block, with the output

$$y^i = F^i(x) + x \tag{10}$$

Then the final output of this 3-level residual group can be expressed as

$$\begin{aligned} y &= \text{Compose}_{i=1}^{3}(y^i) \\ &= \text{Compose}_{i=1}^{3}(F^i(x) + x) \end{aligned} \tag{11}$$

## 3.4 Multiscale residual classifier

Extracting features in traffic is the primary issue of network traffic anomaly detection technology. Network traffic data show different behavior patterns in different time scales and frequencies. Compared with extracting features only from the finest-grained traffic sequence, fully considering multi-scale information can provide additional underlying patterns in the original traffic in a wider range, which has a positive impact on anomaly classification[21]. Therefore, a multi-scale residual classifier (MSRC) is designed. The traffic data is divided into several subsequences by sliding window, and each subsequence is decomposed and reconstructed by wavelet transform technology to generate reconstructed sequence data at different levels. Then the SAE is used to extract and transform the features of each reconstructed sequence to generate new feature data (reconstruction error vector), and input to the residual network constructed by multiple convolution structures. The feature extraction of sequence data at different scales can improve the representation ability of the model. Finally, the results are aggregated and sent to the classifier to get the detection results. Its structure is shown in Figure (6).

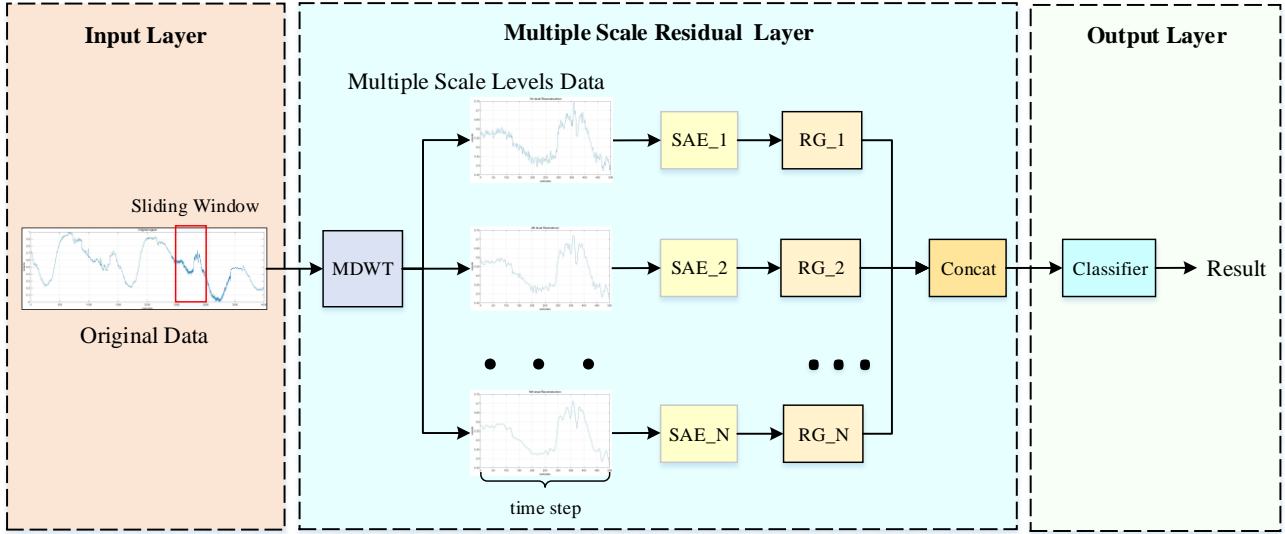

Fig.6: The framework of Multiple Residual Classifiers

For a given number of flows X, a sliding window of scale $m$ is used to divide and using wavelet transform, the reconstructed data subsequence $x^{m,j}$ is obtained according to formula (3), where $j$ is the output subsequence data reconstructed at the $j$th level.

Input it to the corresponding SAE_j, using the reconstructed error vector $e^{m,j}$ that can be obtained for $x^{m,j}$, henceforth referred to as $e$.

In RG, the reconstruction error vector $e$ is mirrored into $n$ different residual blocks, now $R^i$ is the expected transformation of $e$ in the $i$th residual block, and $F^i$ is the residual function, which can be obtained from formula (9)

$$R^i(e) = F^i(e) + e \quad (12)$$

Therefore, the residual group summarizes the overall output of all residual blocks as follows:

$$E = \text{Compose}_{i=1}^n (F^i(e) + e) \quad (13)$$

$\text{Compose}_{i=1}^n$ represents the linear splicing of $n$ residual block link feature maps.

The output layer, which feeds the summarized results of each residual group to the lightweight classifier to predict the results, and outputs the generated prediction labels.

$$y = Classifier(Concat_{j=1}^m (E_j)) \quad (14)$$

where $E_j$ is the output of the $j$th residual group, $m$ is the number of residual groups, $Concat_{j=1}^m$ denotes the output results of the summarized residual groups, and $Classifier(\cdot)$ is the function to achieve the classification prediction.

## IV EXPERIMENTAL SETUP

### 4.1 Experimental environment

The experiments involved in this study were carried out on GPU-supported devices. The GPU model was GeForce GTX 3090Ti, with 24 GB of RAM.

### 4.2 Data preparation

(1) Datasets

Aiming to evaluate the performance of TA-WGAN, we conducted evaluation tests on several network traffic datasets, using four publicly available network traffic datasets including KDD99, NSL-KDD, UNSW-NB15, and CICIDS2018, which were divided into training set, and test set. To reduce computing time, some of their respective data (subsets) are used as the raw data for our study. Table 1 summarizes the basic information of each dataset we use, including the overall number of samples and the number of anomalous samples, the number of features in the traffic and the number of types of attacks that cause anomalies, and each anomalous location in the dataset is known. We directly use the training and test sets already divided by KDD99, NSL-KDD, and UNSW-NB15, and for CICIDS2018, we use the data in Thursday-01-03-2018, and randomly select the normal and abnormal samples in proportion to each other as the training and test sets. Since each dataset has its own characteristics, it makes our anomaly detection work more challenging and also helps us to confirm the validity and limitations of the model.

**TABLE 1:** Network traffic dataset

| Dataset | Dataset-sub | Total | Normal | Anomaly | Features |
|---|---|---|---|---|---|
| KDD99 | 10_percent_corrected | 494021 | 97278 | 396743 | 41 |
| KDD99 | corrected | 253727 | 26053 | 227674 | 41 |
| NSL-KDD | NSL-KDD-Train+ | 25191 | 13448 | 11743 | 41 |
| NSL-KDD | NSL-KDD-Test+ | 22543 | 9711 | 12832 | 41 |
| UNSW-NB15 | NB15_training-set | 82332 | 37000 | 45332 | 49 |
| UNSW-NB15 | NB15_testing-set | 175341 | 56000 | 119341 | 49 |
| CIC-IDS2018 | CIC-IDS2018-train | 198675 | 142822 | 55853 | 79 |
| CIC-IDS2018 | CIC-IDS2018-test | 132425 | 95215 | 37210 | 79 |

Figure 7 shows us visually the ratios of normal and abnormal samples in the four datasets. It is clear that only NSL-KDD and UNSW-NB15 are roughly balanced in the training data (left), while only NSL-KDD is relatively balanced in the test data (right), and several other datasets have large imbalances in both the training and test datasets.

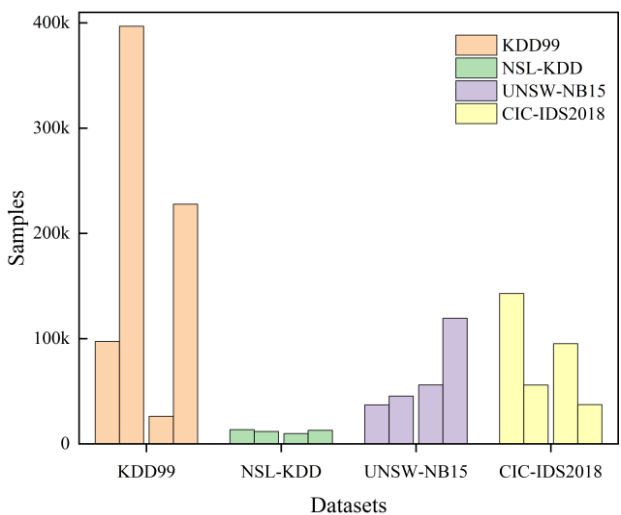

Fig.7: Distribution of normal and abnormal samples for each dataset

(2) Data preprocessing

Data preprocessing is the operation of flow data cleaning, text numerical, data slicing, and value normalization to ensure the readability and uniformity of data

Data cleansing. Traffic data crawled from the real network environment may have duplicate or mutilated invalid data, and these redundant and missing data need to be removed by using data cleaning techniques.

Text numerical. The attribute feature values of the original traffic data are not entirely numerical, but also partly textual information, so it is necessary to convert these texts into corresponding discrete numerical values in order to participate in the operation.

Traffic anonymization. Information such as IP address and MAC address specific to each traffic flow may affect the classification feature extraction. To eliminate the influence of these factors, the original address is replaced with a new randomly generated address. This step is optional and is not required if the traffic to be detected is from the same network environment.

Numerical normalization. Different attributes have different scales, and the range of values of the feature vectors varies, which can affect the detection results when the differences are large.

### 4.3 Evaluation Metrics

To evaluate and validate the performance of the proposed MSRC method, we use $Accuracy$, $Precision$, $Recall$, and $F1$ values, where positive cases represent abnormal traffic and negative cases represent normal traffic. In addition, because of the imbalance in the dataset, we also use $AUC$ (Area Under Curve) as an evaluation metric, which is the area under the $ROC$ curve. The $ROC$ (Receiver Operating Characteristic) curve is based on a series of different cut-off values or thresholds, with the true positive rate ($TRP$) or recall rate ($Recall$) as the vertical coordinate and the false positive rate ($FRP$) as the horizontal coordinate, and the $AUC$ describes the difference between the two types of predictive distributions [22]. Intuitively, the closer the $AUC$ is to 1, the better the classification design is.

These metrics were calculated according to the following formula

$$Accuracy = \frac{TP+TN}{TP+TN+FP+FN}$$

$$Precision = \frac{TP}{TP+FP}$$

$$Recall = TPR = \frac{TP}{TP+FN}$$

$$F_1 = \frac{2Precision*Recall}{Precision+Recall}$$

$$FPR = \frac{FP}{FP+TN}$$

Where $TP$、$TN$、$FP$、$FN$ refer to elements in confusion matrix of Table 2。

**TABLE 2: Confusion Matrix**

|  | Predictive Positive | Predictive Negative |
|---|---|---|
| **True Positive** | TP | FN |
| **True Negative** | FP | TN |

## V  EXPERIMENT

### 5.1  Experiment 1

In this section, we use the DB3 wavelet filter, the sliding window size is 800, the wavelet decomposition scale is 6, the SAE structure is composed of three layers of AE superposition, the input and output dimensions are shown in Table 3, and the residual group consists of three residuals block composition.

**TABLE 3: Architecture of SAE**

| Autoencoder | Size of input | Size of output |
|---|---|---|
| Encoder_1 | 150 | 110 |
| Encoder_2 | 110 | 90 |
| Encoder_3 | 90 | 64 |
| Decoder_1 | 64 | 90 |
| Decoder_2 | 90 | 110 |
| Decoder_3 | 110 | 150 |

To test the fitting ability and generalization ability of our proposed method, MSRC used a ten-fold cross-validation method after completing training on the NSL-KDD dataset, that is, 10% of the data from the training and test sets were taken for testing each time, and the final results were averaged over 10 operations, as detailed in Table 4, and two conclusions can be obtained from the testing results.

**TABLE 4: Results of Experimental**

| Each | NSL-KDD Train+ |  | NSL-KDD Test+(10%) |  |
|---|---|---|---|---|
|  | Accuracy | AU-ROC | Accuracy | AU-ROC |
| 1 | 0.8785 | 0.8995 | **0.8506** | 0.9074 |
| 2 | 0.8778 | 0.9004 | 0.8278 | 0.8513 |
| 3 | **0.8914** | 0.9006 | 0.8301 | 0.9045 |
| 4 | 0.8878 | 0.8994 | 0.8481 | 0.8967 |
| 5 | 0.8762 | 0.8952 | 0.8426 | 0.8815 |
| 6 | 0.8757 | 0.8997 | 0.8296 | 0.8588 |
| 7 | 0.8829 | 0.9004 | 0.8228 | **0.9237** |
| 8 | 0.8883 | 0.9006 | 0.8292 | 0.8831 |
| 9 | 0.8763 | **0.9017** | 0.8219 | 0.8759 |
| 10 | 0.8794 | 0.9014 | 0.8333 | 0.9175 |
| Average | 0.8814 | 0.8999 | 0.8336 | 0.8900 |

(1) Our proposed method can detect anomalous data in network traffic, and the detection accuracy is above 85% in both training and test sets.

(2) The comparative performance of MSRC on the training and test sets shows that the detection accuracy of MSRC on the training set is 88.14% better than that on the test set at 85.46%, indicating that the fitting ability of MSRC is better than the generalization ability; and the AUC-ROC on the two data sets are almost the same, indicating that the model generally performs better regardless of the basic probability of class.

### 5.2  Experiment 2

To validate the performance of our proposed method, we compare it with some typical anomaly detection methods on different data sets, which include:

**TABLE 5: Anomaly Detection Metrics for Different Datasets**

| Model | Dataset | Precision | Recall | F1 |
|---|---|---|---|---|
| RNN | KDD99 | 0.5420 | 0.5311 | 0.5365 |
|  | NSL-KDD | 0.6242 | 0.5564 | 0.5884 |
|  | UNSW-NB15 | 0.6754 | 0.8127 | 0.7377 |
|  | CIC-IDS2018 | 0.5863 | 0.6745 | 0.6273 |

| Model | Dataset | Precision | Recall | F1 |
|---|---|---|---|---|
| | Total | 0.6070 | 0.6437 | 0.6225 |
| | KDD99 | 0.6667 | 0.6832 | 0.6748 |
| | NSL-KDD | 0.6964 | 0.5684 | 0.6259 |
| LSTM | UNSW-NB15 | 0.7803 | 0.8122 | 0.7960 |
| | CIC-IDS2018 | 0.7966 | 0.6803 | 0.7339 |
| | Total | 0.7350 | 0.6860 | 0.7076 |
| | KDD99 | 0.8642 | **0.8811** | **0.8726** |
| | NSL-KDD | 0.8423 | 0.8564 | 0.8493 |
| DAGMM | UNSW-NB15 | 0.8537 | 0.8269 | 0.8401 |
| | CIC-IDS2018 | **0.8632** | 0.7453 | 0.7999 |
| | Total | 0.8559 | 0.8274 | 0.8405 |
| | KDD99 | 0.5367 | 0.6483 | 0.5872 |
| | NSL-KDD | 0.6638 | 0.5837 | 0.6212 |
| MAD-GAN | UNSW-NB15 | 0.8032 | 0.8223 | 0.8126 |
| | CIC-IDS2018 | 0.7659 | 0.8031 | 0.7841 |
| | Total | 0.6924 | 0.7144 | 0.7013 |
| | KDD99 | 0.7875 | 0.7686 | 0.7779 |
| | NSL-KDD | 0.8243 | **0.8810** | 0.8517 |
| Tad-GAN | UNSW-NB15 | 0.8691 | 0.8791 | 0.8741 |
| | CIC-IDS2018 | 0.8383 | 0.8021 | 0.8198 |
| | Total | 0.8298 | 0.8327 | 0.8309 |
| | KDD99 | 0.8379 | 0.8277 | 0.8327 |
| | NSL-KDD | 0.8843 | 0.8610 | 0.8725 |
| CBR-CNN | UNSW-NB15 | 0.8869 | 0.8988 | 0.8928 |
| | CIC-IDS2018 | 0.7684 | 0.8080 | 0.7877 |
| | Total | 0.8444 | 0.8489 | 0.8464 |
| | KDD99 | **0.8845** | 0.8362 | 0.8597 |
| | NSL-KDD | **0.8943** | 0.8705 | **0.8822** |
| MSRC | UNSW-NB15 | **0.9212** | **0.9089** | **0.9150** |
| | CIC-IDS2018 | 0.8088 | **0.8782** | **0.8421** |
| | Total | **0.8772** | **0.8734** | **0.8747** |

(1) RNN and LSTM[23], which are two classical deep neural networks widely used in time series analysis, and both networks use a single latent layer structure in this case.

(2) DAGMM[24] combines deep self-coding compression network and improved Gaussian mixture model (GMM) for anomaly detection means, which had achieved good results on KDD99 dataset.

(3) MAD-GAN and Tad-GAN, which are classical methods for time series for anomaly detection using generative adversarial networks, have performed well in anomaly detection experiments on time series.

(4) CBR-CNN, a deep convolutional neural network (CBR-CNN) structure built using channel enhancement and residual learning for detecting network intrusions, which uses multiple SAEs for multiplex mapping of the original signal to achieve channel enhancement, and then uses residual networks to learn different granularity features of each signal.

Each model is run 10 times on each dataset to obtain the performance including *Precision*, *Recall*, and *F1* value, and use their average classification as a reference to evaluate the model performance.

Table 5 shows us the precision, recall and F1 values of the six models for anomaly detection on the four datasets. It can be found that our proposed multiscale residual feature anomaly detection method (MSRC) has the highest *Precision*, *Recall* and *F1* value on the UNSW-NB15 dataset, the best detection precision on the 2 datasets KDD99 and NSL-KDD99, and the highest overall rating of *Precision*, *Recall* and *F1* value on the 4 datasets, which can be visualized in Fig. 8. The excellent performance of MSRC on different datasets indicates that MSRC can be trained and tested across datasets and has strong generalization ability. Although these four datasets come from different network environments and have different number of features and types of attacks, they are all generated from computer networks in a broad sense, and the network traffic data have some commonalities.

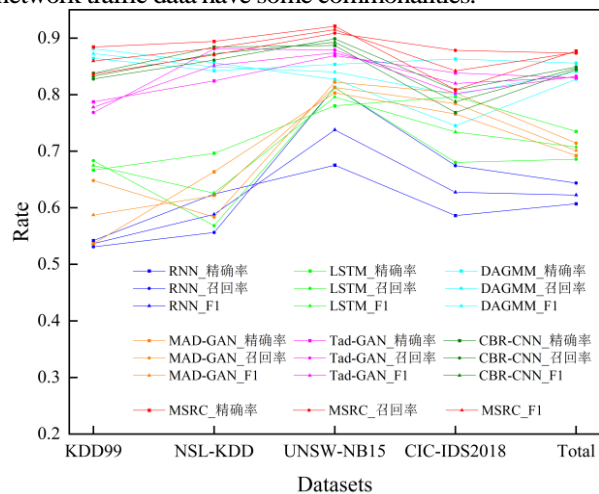

Fig.8  Anomaly Detection Metrics

DAGMM performs better on the KDD99 and NSL-KDD99 datasets, but does not perform well on the UNSW-NB15 and CICIDS2018 datasets, probably because DAGMM itself is a model that has been optimized and adjusted many times on the KDD99 network traffic dataset, so it is more adaptable to the KDD series datasets, while for features different from Tad-GAN and MAD-GAN are both detection methods with the basic structure of generative adversarial networks, and Tad-GAN is obviously better than MAD-GAN in performance, because Tad-GAN uses a double-loop GAN structure to better capture the potential information in the traffic, which is conducive to the identification of abnormal data; while MAD-GAN was originally designed to detect anomalous data in water treatment and distribution systems, and is not good at anomaly discovery in network traffic. This also shows that there is no one algorithm that can solve all problems, and the algorithms should be different as long as the data sets are different [25].

The overall performance of traditional RNN and LSTM is poor, but better than MAD-GAN. Because network traffic is also a kind of time series data with backward and forward correlation, the recurrent neural network with memory effect is good at extracting feature information in time series (network traffic), and LSTM increases "gate" compared with ordinary RNN. The design of "gate" relieves the long time dependence of ordinary RNN, so LSTM achieves better results.

In addition, it can be found through the curves that, except for DAGMM, the performance indexes of the other five models on the UNSW-NB15 dataset are better than those of the other three datasets. The analysis may be due to the fact that the categories in the UNSW-NB15 training set are more balanced, which can make the model fully trained, while the number of abnormal samples in the test set is almost twice the number of normal samples, and this sample category This imbalance of sample categories brings convenience to the detection task with the purpose of finding abnormalities.

### 5.3 Experiment 3

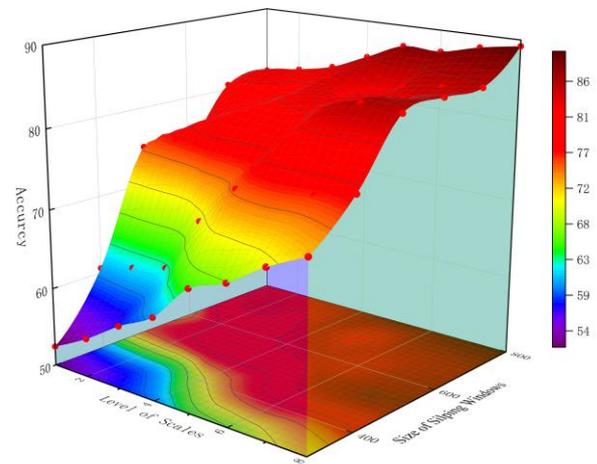

Fig.9   Accuracy of different scale levels and sliding windows

To evaluate the effect of time and frequency scales on the detection results, the detection *Accuracy* on the test data of NSL-KDD was obtained using DB3 wavelet filters at different transform levels, and using different sizes of sliding window sizes on the training set of the dataset NSL-KDD after completing the training, still using the ten-fold cross-validation. From Figure 9, we can see that although there are local concave phenomena on the surface, the overall trend is that the detection effect of subseries with large time scale is better than that of small time scale, which is because the larger the value of sliding window size is used, the richer the information is and the higher the accuracy of detection is; in addition, the more layers of network traffic are decomposed and reconstructed, the better the effect of anomaly detection is, which is consistent with our previous prediction, the reason This is consistent with our previous prediction, because the deeper the layers of wavelet decomposition and reconstruction, the more the model can capture the most essential feature information of the original traffic, and also obtain the diversity information and finer time granularity in different layers, so increasing the number of transform layers can also significantly improve the detection performance of the model.

## VI CONCLUSION

In this study, our goal is to provide diversity information for anomalous traffic detection by analyzing the characteristics of network traffic at different time and frequency scales. To this end, a multi-scale residual feature based on wavelet transform and residual learning is proposed as a network traffic anomaly detection method (MSRC) for identifying and detecting anomalous traffic in the network. Using sliding window and wavelet transform techniques, the feature information of network traffic at different time and frequency scales is captured; the SAE network is used to construct the similarity space, and the reconstruction of the original samples is completed in the similarity space, and the reconstruction error vector is calculated by comparing the differences between the reconstructed samples and the original samples; the multi-path residual network group is used to the error vector, and the features are learned and transformed at different granularity levels, and output the transformed features, and the output layer completes the summary of feature results on each scale and gives the classification prediction results.

The results of Experiment 1 show that our proposed MSRC anomaly detection method, which preserves the feature information in the original traffic after wavelet decomposition and reconstruction of the original traffic data, can be used to be able to accomplish the task of identifying anomalies in the original traffic. In Experiment 2, the proposed method is verified to have high detection performance and good generalization ability to data with different characteristics by comparing with several typical anomaly detection models. The results in Experiment 3 show that large time span and more decomposition levels have a positive effect on mining the diversity information in time correlated data.